\documentclass[preprint,spanish,aps,ams,showpacs,superscriptaddress]{revtex4}

\usepackage{graphicx,color}
\usepackage{latexsym}
\usepackage{amsmath,amssymb}        
\usepackage[draft=false]{hyperref}
\usepackage{mathrsfs}

\usepackage{mathrsfs}
\DeclareSymbolFontAlphabet{\mathrsfs}{rsfs}
\DeclareMathAlphabet{\mathcal}{OMS}{cmsy}{m}{n}

\begin{document}


\title{The shocks during the accretion of an ultrarelativistic supersonic gas onto a rotating black hole}


\author{A. Cruz-Osorio, F. D. Lora-Clavijo and F. S. Guzm\'an}
\affiliation{Instituto de F\'{\i}sica y Matem\'{a}ticas, Universidad
              Michoacana de San Nicol\'as de Hidalgo. Edificio C-3, Cd.
              Universitaria, 58040 Morelia, Michoac\'{a}n,
              M\'{e}xico.}


\date{\today}


\begin{abstract}
In this work, we track the evolution of an ultrarelativistic fluid onto a Kerr black hole, on the equatorial plane. In this treatment, we consider the limit where the rest mass density is neglected, that is, the approximation is valid in the regime where the internal energy dominates over the rest mass density. We particularly concentrate in the case of a gas with $\Gamma=4/3$, which corresponds to a radiation fluid. We show, as in several cases, that a shock cone appears when the asymptotic velocity of the fluid is larger than the asymptotic relativistic sound speed of the gas. On the other hand, in order to show the system approaches to steady state, we calculate the accreted total energy rate on a spherical surface. Finally, we also show the gas distribution and various of its properties.
\end{abstract}




\maketitle


\section{Introduction}

Bondi-Hoyle accretion is the evolution of a homogeneously distributed gas moving uniformly toward a central compact object \cite{Hoyle}. A summary of studies in Newtonian gravity can be found in \cite{Foglizzo}. The relativistic Bondy-Hoyle accretion was first analyzed in \cite{Petrich}, where axially symmetric numerical simulations were carried out in order to study the different matter patterns developed by the gas during the accretion process  onto a black hole. Later, the authors in \cite{Fontnoaxi1,Fontaxi,Fontnoaxi2} studied the relativistic Bondi-Hoyle accretion onto Schwarzschild  and  Kerr black holes, in one case considering axisymmetric fluxes and in a second case, general non-axisymmetric fluxes in the equatorial plane using s-lab symmetry, that is, whatever happens on the equatorial plane is assumed to happen along the directions perpendicular to it. The authors confirm the formation of a shock cone when the fluid is supersonic and study the dependence of the morphology on the parameters of the wind and the black hole. In astrophysical contexts, in \cite{Donmez} it was shown that the shock cone vibrations can be associated with sources of high energy Quasi Periodical Oscillating signals (QPOs), and found a flip-flop type of unstable oscillation of the shock cone. However later, it was shown in \cite{CruzLora2012} that the flip-flop oscillation of the shock cone depends on the coordinates used to describe the rotating black hole, specifically it was found that the flip-flop oscillation does not appear when penetrating coordinates are used. On the other hand, considering axially symmetric fluxes, the shock cone oscillation, as a potential source of QPOs were studied in \cite{Lora}. More realistic scenarios incorporate astrophysically relevant ingredients like magnetic fields \cite{Penner2} and radiative terms \cite{Zanotti} to the Bondi-Hoyle accretion.

\noindent The ultrarelativistic Bondi-Hoyle accretion on a rotating black hole was recently reported in  \cite{Penner}, considering axisymmetric fluxes. In this work we present a numerical study of ultrarelativistic Bondi-Hoyle accretion on a rotating black hole in the equatorial plane with s-lab symmetry. In our study we describe the rotating black hole space-time using Kerr-Schild coordinates. Specifically, we only focus in supersonic fluxes.


\section{Ultrarelativistic Hydrodynamics Equations}
\label{sec:equations}

We use a perfect fluid to model the ultrarelativistic gas, with  stress-energy tensor $T^{\mu \nu} =( \rho + p) u^{\mu}u^{\nu} + pg^{\mu\nu}$, where  $u^{\mu}$ are the components of the 4-velocity of a fluid element, $p$ its pressure and $g_{\mu\nu}$ are the components of the metric of the space-time background. The ultrarelativistic Euler equations can be derived from the local conservation of the stress-energy tensor $\nabla_{\mu}T^{\mu\nu} = 0$, where $\nabla_{\mu}$ is the covariant derivative consistent with the 4-metric $g_{\mu\nu}$. The conservation of mass is satisfied identically in the ultrarelativistic case.

\noindent An ultrarelativistic fluid has internal energy sufficiently large that the rest mass density is negligible, that is $\epsilon\rho_{0} \gg \rho_{0}$, where $\epsilon$ is the specific internal energy. This condition implies that approximately the energy density is $\rho=\rho_{0}+\epsilon\rho_{0} \sim \epsilon\rho_{0}$ \cite{Choptuik}. We use an ideal gas equation of state $p=(\Gamma-1)\rho_0 \epsilon$ which reduces to $p = (\Gamma-1)\rho$ for the ultrarelativistic case, where $\Gamma$ is the ratio between specific heats, that corresponds for example, to radiation when $\Gamma=4/3$.

\noindent We use the standard 3+1 decomposition to describe the space-time background 
\begin{equation}
ds^{2}= -\alpha^{2}dt^{2} + \gamma_{ij}(dx^{i}+\beta^i dt)(dx^{j} + \beta^j dt),
\end{equation}
\noindent where $\beta^i$ is the shift vector, $\alpha$ the lapse function and $\gamma_{ij}$ are the components of the spatial 3-metric. Using this decomposition of space-time, the ultrarelativistic Euler equations, on the equatorial plane  ($\theta=\pi/2$), can be written in flux balance law form \cite{banyuls} as

\begin{eqnarray}
\partial _{t} {\bf u} + \partial _{r} \big(\alpha {\bf F}^{r}  \big)  +\partial _{\phi} \big(\alpha {\bf F}^{\phi}  \big) 
 = \alpha {\bf S} - \frac{\partial_{r} \sqrt{\gamma}}{\sqrt{\gamma}}  \big(\alpha {\bf F}^{r}  \big) , \label{eq:flux_conservative}
\end{eqnarray}

\noindent where $\gamma=det(\gamma_{ij})$ is the determinant of the spatial 3-metric, ${\bf u}$ is the vector whose entries are the conservative variables, which are in terms of the original primitive variables ${\bf W} = (v^{i},p,\epsilon)$. The vectors ${\bf F}^{r}$ and ${\bf F}^{\phi}$ are the fluxes along the $r$ and $\phi$ spatial directions respectively and ${\bf S}$ is a source vector field. Specifically, the fields in (\ref{eq:flux_conservative}) are

\begin{eqnarray} 
{\bf u}&=& \left[ S_{r}, S_{\phi}, \tau )\right] = \left[ (\rho + p) W^{2}v_{r} , (\rho + p) W^{2}v_{\phi}, (\rho + p) W^{2} - p \right], \nonumber \\ 
{\bf F}^{r} &=& \left[ \Big(v^{r}-\frac{\beta^{r}}{\alpha}\Big)S_{r} + p, \Big(v^{r}-\frac{\beta^{r}}{\alpha}\Big)S_{\phi}, \Big(v^{r}-\frac{\beta^{r}}{\alpha}\Big) \tau + pv^{r} \right],\\ 
{\bf F}^{\phi} &=& \left[ v^{\phi} S_{r}, v^{\phi} S_{\phi}+ p, v^{\phi} \big(\tau+p\big) \right], \nonumber\\
{\bf S} &=&\left[ T^{\mu \nu} g_{\nu \sigma } \Gamma ^{\sigma}_{\mu r}, 0, T^{r t} \partial_{r}\alpha - T^{\mu \nu} \Gamma^{t}_{\mu\nu} \alpha \right], \nonumber
\end{eqnarray}

\noindent where $v^i$ are the components of the 3-velocity of the gas measured by an Eulerian observer, which are related to the spatial components of the 4-velocity of the fluid elements by $v^{i}=u^{i}/W + \beta^{i}/\alpha$, $W=1/\sqrt{1-\gamma_{ij}v^{i}v^{j} }$ is the Lorentz factor and $\Gamma^{\sigma}_{\alpha\beta} $ are the Christoffel symbols of the space-time metric.

\section{Numerical Methods}
\label{sec:numerics}

{\it Domain and boundary conditions.} We study numerically the ultrarelativistic gas on the equatorial plane, in the domain $ [r_{exc},r_{max}] \times  [0,2\pi)$. We choose the interior boundary $r_{exc}$ to be inside the black hole horizon, were we apply excision \cite{SeidelSuen1992}. The exterior boundary $r_{max}$, is splitted in to two halves, one in which the gas enters the domain where we apply inflow boundary conditions, and a second half where the gas leaves the domain and we apply outflow boundary conditions there.
  
{\it Evolution.} We use a high resolution shock capturing method, using the HLLE approximate Riemann solver formula \cite{hlle}, and the minmod variable reconstructor. We evolve in time with a second order Runge Kutta integrator \cite{Shu}. The numerical grid is uniformly spaced with resolutions $dr,d\phi$. We use a constant time step given by $d t=C {\it min}(dr,d\phi)$, where $C=0.25$ is a constant in time and space Courant factor estimated empirically to maintain stability. 

{\it Initial Data.} We center the black hole at the origin of the coordinates origin. We consider a homogeneous ultrarelativistic gas, that uniformly fills the whole domain, moving on the equatorial plane along the $x$ direction with a constant density initially. In the ultrarelativistic case, the speed of sound depends only on $\Gamma$, $c_{s}= \sqrt{\Gamma -1}$. Then we choose the initial pressure $p_{0}$ to be a constant. The initial velocity field $v^i$ can be expressed in terms of the asymptotic velocity $v_{\infty}$, as done for relativistic winds \cite{Fontnoaxi2}, where the relation $v^2=v_iv^i=v^2_{\infty}$ is satisfied. We set the  black hole mass $M=1$, and the adiabatic index $\Gamma=4/3$. We choose the gas velocity initially $v_{\infty}=0.9$ and explore four values of the black hole's angular momentum $a=0,0.5,0.7,0.9$.

{\it Recovery of the primitive variables.} We recover the primitive variables $(v^{i},p)$ exactly during the evolution. The pressure and components of the velocity in terms of the conservative variables are respectively $p=-2\sigma \tau +\sqrt{ 4\sigma^{2}\tau^{2} +  (\Gamma -1)(\tau^{2}-S^{2}) }$, $v^{i}=\frac{S_{i}}{\tau+p}$, where $\sigma =\frac{2-\Gamma }{4}$ \cite{Choptuik}. From this we can reconstruct the energy density $\rho$ and the Lorentz factor $W$. We also use an atmosphere to avoid the divergence of the internal enthalpy. 

Standard numerical tests of our implementation for the general case of relativistic hydrodynamics can be found in \cite{Lora}.


\section{Results}
\label{sec:results}
\begin{figure}
\includegraphics[width=10cm]{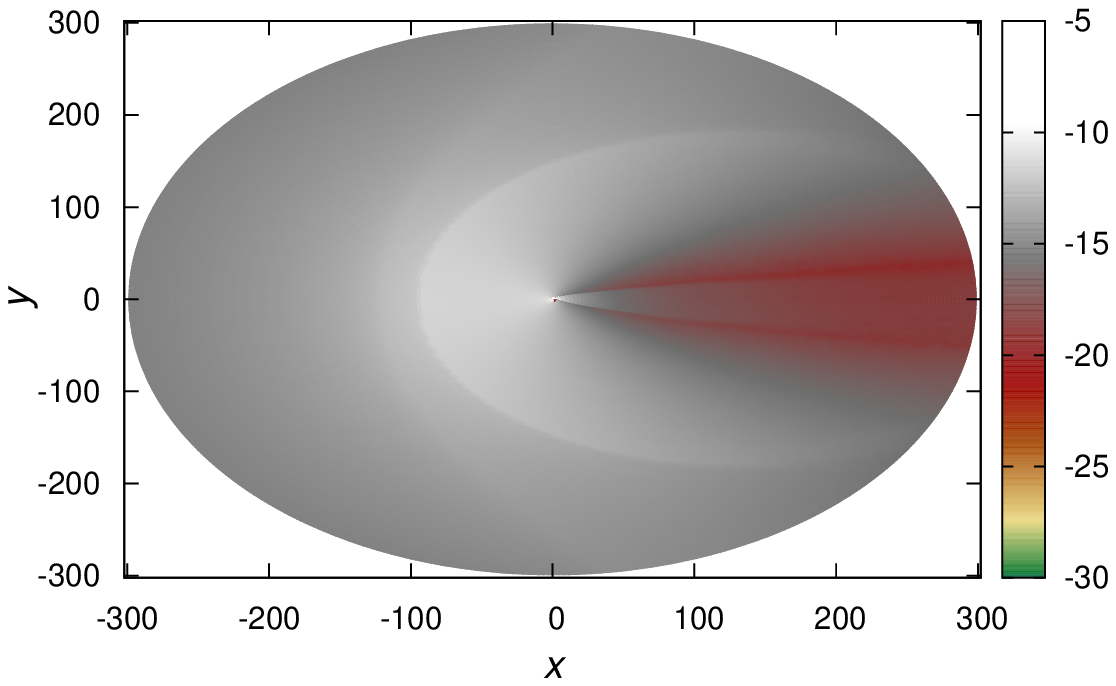}
\includegraphics[width=10cm]{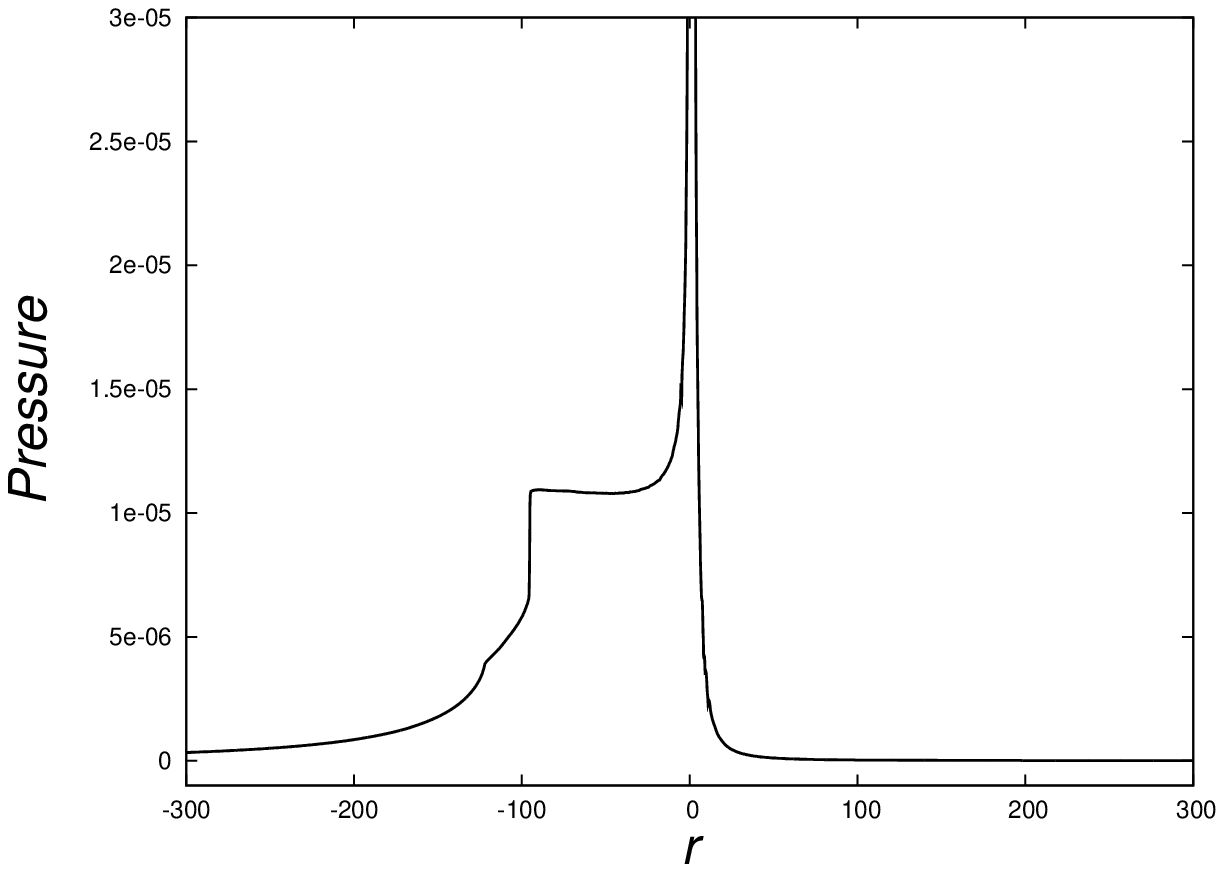}
\caption{ On the top panel, we show the morphology of the pressure profile of the gas for the case of $v_{\infty}=0.9$ and $a=0.9$. We appreciate the shock cone in the dark region (red in color) and the bow shock in light color. On the bottom panel, we show a one dimensional slice of the pressure profile along the $x$ axis. This shows a high pressure zone in the bow shock region compared to the low pressure in shock cone.}
\label{fig:1}
\end{figure}

\begin{figure}
\includegraphics[width=8cm]{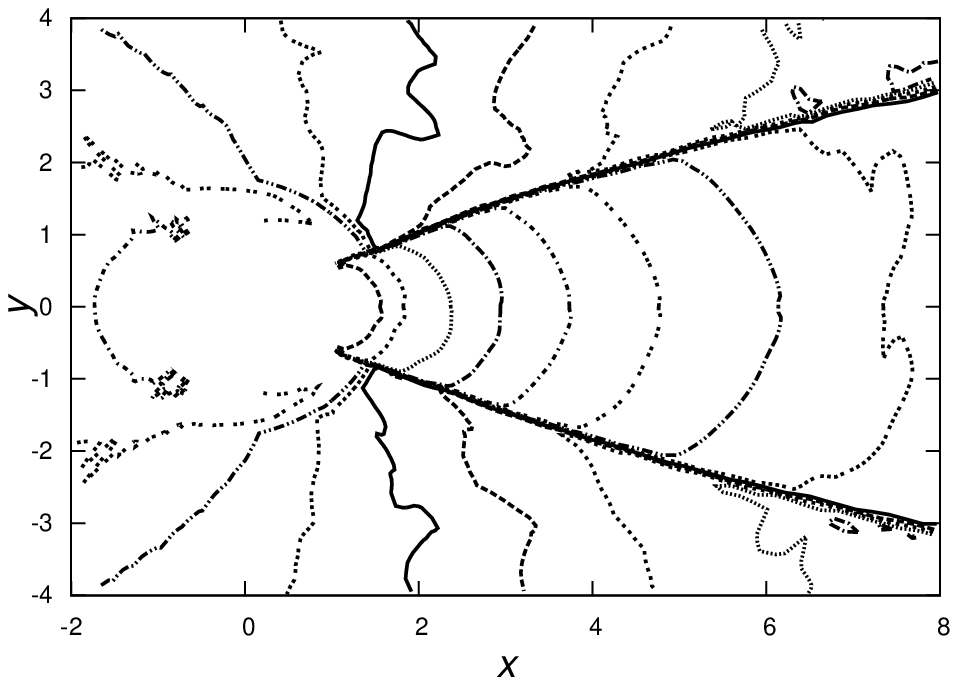}
\includegraphics[width=8cm]{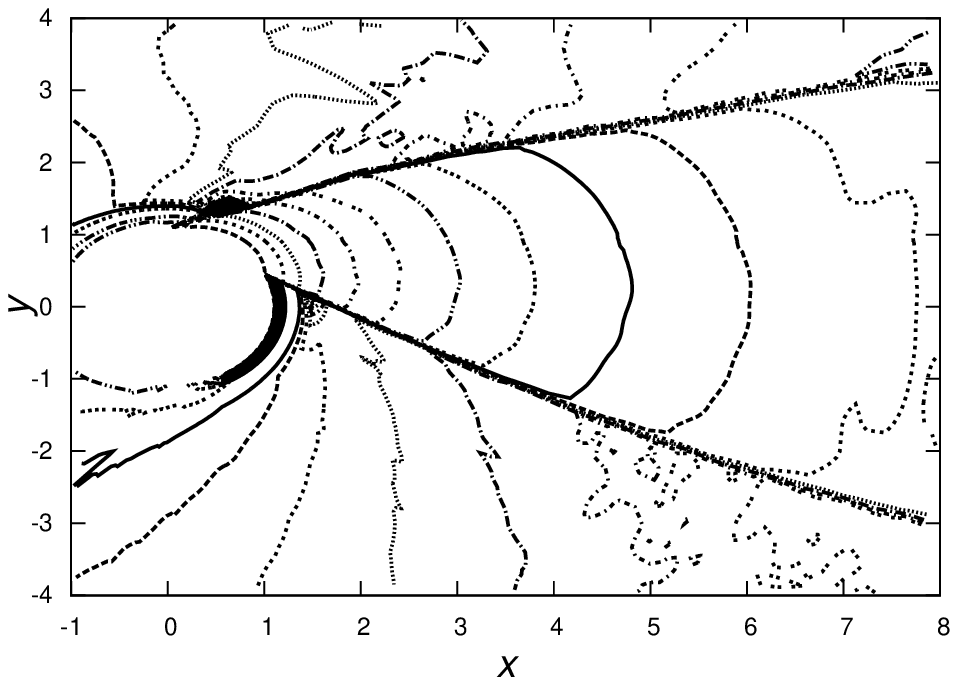}
\caption{We show the isocontours of the pressure. We can appreciate the shock cone deformation by the effects of the rotation of the back hole. We show the morphology of the fluid when the angular momentum of the black hole takes the values $a=0$ (left) and $a=0.9$ (right) respectively.}
\label{fig:2}
\end{figure}

{\it Morphology.} The ultrarelativistic gas that is moving with supersonic velocity is characterized by the formation of the shock cone in the rear part of the black hole, which is a low pressure region and a bow shock at the front side of the hole, which is a high pressure zone expanding spherically in opposite direction to the gas flow.
In Fig. \ref{fig:1} we show the morphology of the pressure for a gas with a supersonic velocity $v_{\infty}=0.9$, feeding a rotating black hole with angular momentum $a=0.9$. 
We can see that the shock cone is formed (the dark zone in the left panel). At the same time, a bow shock appears expanding, shown in light color in the plot. On the other hand, in the right panel we show a slice of the pressure along the $x$ axis at a given time, in which we can appreciate the high pressure in the bow shock and a low pressure in shock cone. We found a similar behavior of the pressure for different values of the black hole angular momentum. 
In Fig. \ref{fig:2} we show an example of how the rotation of the black hole deforms the shock cone near the event horizon. If the effects of the rotation of a black hole accreting ultrarelativistic (and relativistic \cite{CruzLora2012}) gas are to be measured, they have to be associated to the regions pretty close to the black hole.

\begin{figure}
\includegraphics[width=8cm]{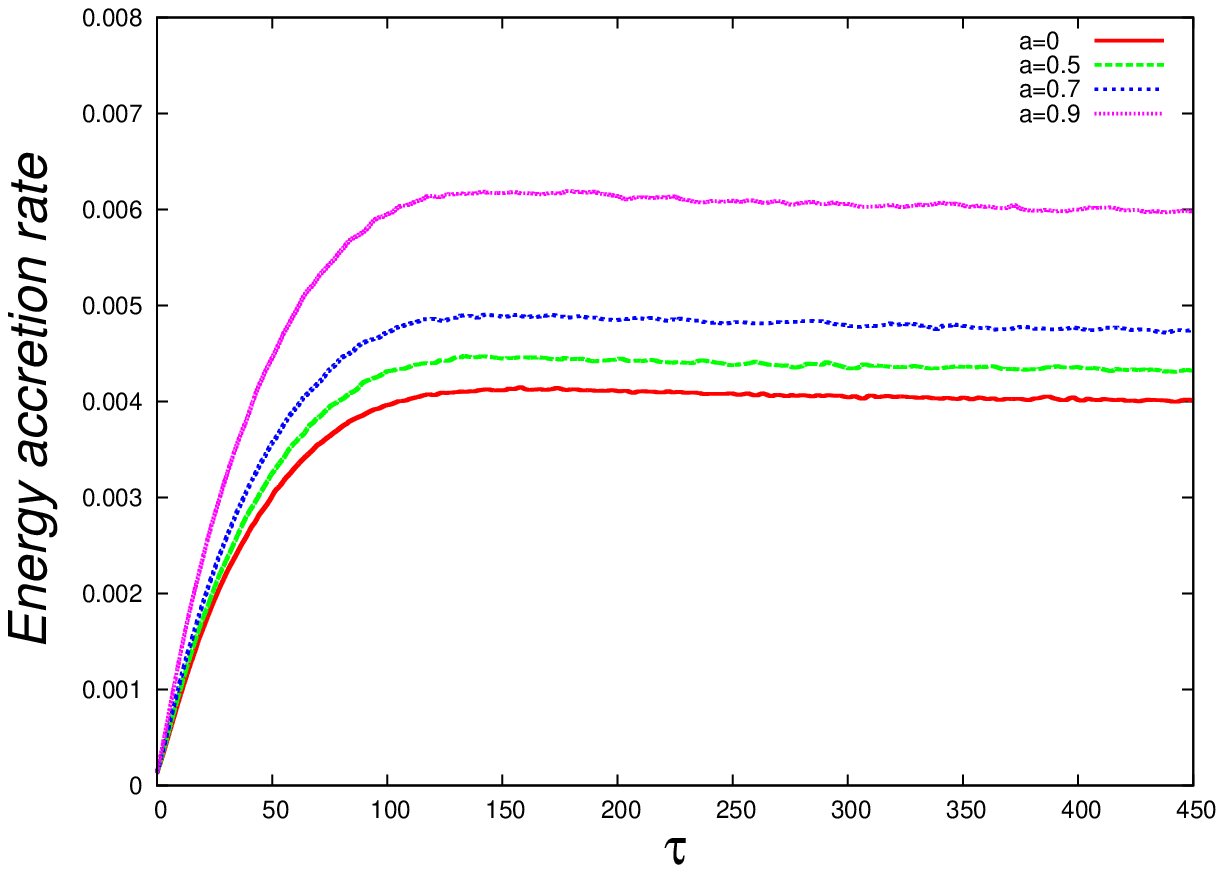}
\caption{ Total energy accretion rate for different values of the  rotation parameter of the black hole. Here we present the energy accretion rate measured in a detector located at $r=2M$, we can see that the accretion rate approaches to stationary regime and that it is bigger for bigger values of the the rotation parameter of the hole.}
\label{fig:3}
\end{figure}

{\it Diagnostics.} We compute the accreted energy rate on spherical a surface. In the equatorial plane the accreted total energy rate $\dot{\cal{E}}$ is

\begin{equation}
\dot{\cal{E}} = \int^{2\pi}_{0}  \alpha\sqrt{\gamma} \Big[ \tau(v^{r}-\beta^{r}) + pv^{r} \Big]d\phi  + \int^{2\pi}_{0} \int^{r_{det}}_{r_{exc}}  \alpha\sqrt{\gamma}  S_{\tau}drd\phi,
\end{equation}

\noindent where $S_{\tau}$ is the source term corresponding to the evolution equation of the conservative variable $\tau$  and $r_{det}$ is the position of the detector where we measure the total energy rate. We placed various detectors at different radii. In Fig. \ref{fig:3} we show the total energy accretion rate in proper time for different values of a rotating black hole, measured at $r=2$. The accretion rate initially grows and approaches to stationary state. We found that the bigger the angular momentum of the black hole the higher the accretion rate.

\section{Conclusions}
\label{sec:conclusions}
We have shown that the rotation of the black hole induces a deformation of the shock cone at short distances from the event horizon and not far from the black hole, where we expect the streaming properties of the gas dominate over the gravitation field. A bow shock is formed independently of the angular momentum of the black hole. We also conclude that when $a$ increases, the total energy accretion rate is higher. 


\section*{Acknowledgments}

This research is partly supported by grants CIC-UMSNH-4.9 and CONACyT 106466.


\end{document}